%% file: llived_prl.tex
\documentclass[aps,prl,twocolumn,superscriptaddress,floatfix,showpacs]{revtex4}
\usepackage{epsfig}
\usepackage{graphicx}
\usepackage{dcolumn}
\usepackage{amsmath}
\usepackage{xspace}
\graphicspath{./}

\begin{document}
\input{phys_defs.tex}


\title{\bf Search for Heavy, Long-Lived Particles that Decay to
  Photons at CDF II}
\preprint{FERMILAB-PUB-06/yyy-E}

\date{\today; Version 5.1}

\input{Jan_2007_Authors_Visitors}

\begin{abstract}
  We present the first search for heavy, long-lived particles that
  decay to photons at a hadron collider. We use a sample of
  $\gamma$+jet+missing transverse energy events in $p{\bar p}$
  collisions at $\sqrt{s}=1.96$~TeV taken with the CDF II detector.
  Candidate events are selected based on the arrival time of the
  photon at the detector. Using an integrated luminosity of
  570~\invpb\ of collision data, we observe 2 events, consistent with
  the background estimate of 1.3$\pm$0.7 events. While our search
  strategy does not rely on model-specific dynamics, we set cross
  section limits in a supersymmetric model with \NTGTGR\ and place the
  world-best 95\% C.L. lower limit on the \none\ mass of 101~\gevc\ at
  $\tau_{\tilde{\chi}_1^0}$~=~5~ns.
\end{abstract}


\maketitle
\newcounter{myctr}

\clearpage

Searches for events with final state photons and missing transverse
energy (\mett)~\cite{rapdef} at collider experiments are sensitive to
new physics from a wide variety of models~\cite{feng} including gauge
mediated supersymmetry breaking (GMSB)~\cite{gmsb}. In these models
the lightest neutralino (\neut) decays into a photon (\photon) and a
weakly interacting, stable gravitino (\grav) that gives rise to \mett\
by leaving the detector without depositing any energy. The observation
of an \eeggmett\ candidate event by the CDF experiment during Run~I at
the Fermilab Tevatron~\cite{eeggmet} has increased the interest in
experimental tests of this class of theories. Most subsequent searches
have focused on promptly produced photons~\cite{ggmet,d0ggmet},
however the \neut\ can have a lifetime on the order of nanoseconds or
more. This is the first search for heavy, long-lived particles that
decay to photons at a hadron collider.

We optimize our selection requirements using a GMSB model with a
standard choice of parameters~\cite{snowmass} and vary the values of
the \neut\ mass and lifetime. However, the final search strategy is
chosen to be sufficiently general and independent of the specific GMSB
model dynamics to yield results that are approximately valid for any
model producing the same reconstructed final state topology and
kinematics~\cite{prospects}. In $p{\bar p}$ collisions at the Tevatron
the inclusive GMSB production cross section is dominated by pair
production of gauginos. The gauginos decay promptly, resulting in a
pair of long-lived \none's in association with other final state
particles that can be identified as jets. For a heavy \none\ decaying
inside the detector, the photon can arrive at the face of the detector
with a time delay relative to promptly produced photons. To have good
sensitivity for nanosecond-lifetime \none's~\cite{prospects}, we
search for events that contain a time-delayed photon, \mett, and
$\ge1$~jet. This is equivalent to requiring that at least one of the
long-lived \none's decays inside the detector.


This Letter summarizes~\cite{pwt} the first search for heavy,
long-lived particles that decay to photons at a hadron collider. The
data comprise 570$\pm$34~\invpb\ of $p{\bar p}$ collisions collected
with the CDF~II detector~\cite{CDFII} at $\sqrt{s}=1.96$~TeV. Previous
searches for nanosecond-lifetime particles using non-timing techniques
yielded null results~\cite{lep}.


A full description of the CDF~II detector can be found
elsewhere~\cite{CDFII}. Here we briefly describe the aspects of the
detector relevant to this analysis. The magnetic spectrometer consists
of tracking devices inside a \mbox{3-m} diameter, 5-m long
superconducting solenoid magnet that operates at 1.4~T. An eight-layer
silicon microstrip detector array and a 3.1-m long drift chamber with
96 layers of sense wires measure the position ($\vec{x}_i$) and time
($t_i$) of the $\pbarp$ interaction~\cite{zvtime} and the momenta of
charged particles. Muons from collisions or cosmic rays are identified
by a system of drift chambers situated outside the calorimeters in the
region with pseudorapidity $|\eta|<1.1$~\cite{rapdef}. The calorimeter
consists of projective towers with electromagnetic and hadronic
compartments. It is divided into a central barrel that surrounds the
solenoid coil (\mbox{$|\eta|<1.1$}) and a pair of end-plugs that cover
the region $1.1<|\eta|<3.6$.  Both calorimeters are used to identify
and measure the energy and position of photons, electrons, jets, and
\mett.  The electromagnetic calorimeters were recently instrumented
with a new system, the EMTiming system (completed in Fall
2004)~\cite{nim}, that measures the arrival time of electrons and
photons in each tower with $|\eta|<2.1$ for all energies above
$\sim$5~GeV.


The time and position of arrival of the photon at the calorimeter,
$t_f$ and $\vec{x}_f$, are used to separate the photons from the
decays of heavy, long-lived \none's from promptly produced photons or
photons from non-collision sources. We define the corrected arrival
time of the photon as
\begin{equation*}
  t_{c}^{\gamma} \equiv t_f - t_i - \frac{|\vec{x}_f - \vec{x}_i|}{c}.
\end{equation*}
The $t_{c}^{\gamma}$ distribution for promptly produced, high energy
photons is Gaussian with a mean of zero by construction and with a
standard deviation that depends only on the measurement resolution
assuming that the $p{\bar p}$ production vertex has been correctly
identified. Photons from heavy, long-lived particles can have arrival
times that are many standard deviations larger than zero.


The analysis preselection is summarized in Table~\ref{tab:select}. It
begins with events passing an online, three-level trigger by having a
photon candidate in the region $|\eta|<1.1$ with \ett$>25$~GeV and
\mett$>25$~GeV. Offline, the highest \ett\ photon candidate in the
fiducial region of the calorimeter is required to have \ett~$>30$~GeV
and to pass the standard photon identification
requirements~\cite{ggmet} with a minor modification~\cite{photons}. We
require the event to have \mett~$>30$~GeV where the trigger is 100\%
efficient. We require at least one jet with
$|\eta^{\mathrm{jet}}|<2.0$ and
$E^{\mathrm{jet}}_{T}>30$~GeV~\cite{jets}. Since a second photon can
be identified as a jet, the analysis is sensitive to signatures where
one or both \none's decay inside the detector. To ensure a high
quality $t_i$ and $\vec{x}_i$ measurement, we require a vertex with at
least 4 tracks, $\sum_{\rm tracks} p_{T}>15$~\punit, and
$|z_i|<$~60~cm; this also helps to reduce non-collision backgrounds.
For events with multiple reconstructed vertices, we pick the vertex
with the highest $\sum_{\rm tracks} p_{T}$. To reduce cosmic ray
background, events are rejected if there are hits in a muon chamber
that are not matched to any track and are within 30$\deg$\ of the
photon.  After the above requirements there are 11,932 events in the
data sample.

\scriptsize
\begin{table}[tbp]
  \begin{center} 
    \begin{tabular}{l@{\hspace{-14.5pt}}c} \hline \hline
      \bf{Preselection Requirements} & \bf{Cumulative (individual)} \\
      & \bf{Efficiency (\%)} \\
      \ $E_{T}^{\gamma} > 30$~GeV, $\met>30$~GeV& 54 (54) \\
      \ Photon ID and fiducial, $|\eta|<1.0$~~~ & \ 39 (74)* \\
      \ Good vertex, $\sum_{\rm tracks}p_{T}>15$~\punit\ \ & 31 (79) \\
      \ $|\eta^{\mathrm{jet}}|<2.0$, $E^{\rm jet}_{T}>30$~GeV & 24 (77) \\
      \ Cosmic ray rejection  & \ 23 (98)* \\
      \bf{Requirements after Optimization} & \\
      \ $\met>40$~GeV, $E^{\rm jet}_{T}>35$~GeV & 21 (92) \\
      \ $\Delta\phi$(\mett, jet) $>$ 1 rad & 18 (86) \\
      \ 2~ns $< t_{c}^{\gamma} <$ 10~ns & 6 (33) \\
      \hline \hline
    \end{tabular}    
    \caption{\label{tab:select} The data selection criteria and the
      cumulative and individual requirement efficiencies for an
      example GMSB model point at $m_{\tilde{\chi}_1^0}$~=~100~\gevc\
      and $\tau_{\tilde{\chi}_1^0}$~=~5~ns. The efficiencies listed
      are, in general, model-dependent and have a fractional
      uncertainty of 10\%. Model-independent efficiencies are
      indicated with an asterisk. The collision fiducial requirement
      of $|z_i|<60$~cm is part of the good vertex requirement (95\%)
      and is estimated from data.}
  \end{center}
\end{table}
\normalsize

\begin{figure}[tbp]
  \centering%
  \includegraphics[width=1\linewidth]{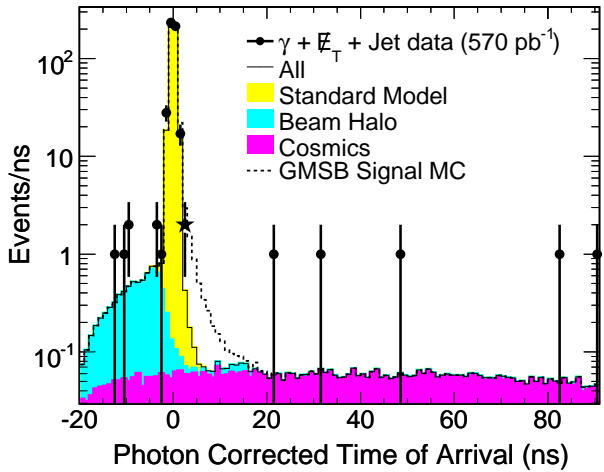}
  \caption{The time distribution for photons passing all but the final
    timing requirement for the background predictions, data, and a
    GMSB signal for an example point at
    $m_{\tilde{\chi}_1^0}$~=~100~\gevc,
    $\tau_{\tilde{\chi}_1^0}$~=~5~ns. A total of 1.3$\pm$0.7
    background events are predicted and 2 (marked with a star) are
    observed in the signal region of $2<t_{c}^{\gamma}<10$~ns.}
  \label{fig:data}
\end{figure}

There are two major classes of background events: collision and
non-collision photon candidates. Collision photons are presumed to
come from standard model interactions, e.g., \photon+jet+mismeasured
\mett, dijet+mismeasured \mett\ where the jet is mis-identified as a
\photon, and \Wele\ where the electron is mis-identified as a \photon.
Non-collision backgrounds come from cosmic rays and beam effects that
can produce photon candidates, \mett, and sometimes the reconstructed
jet. We separate data events as a function of $t_{c}^{\gamma}$ into
several control regions that allow us to estimate the number of
background events in the final signal region by fitting to the data
using collision and non-collision shape templates as shown in
Fig.~\ref{fig:data}.


Collision photons are subdivided in two subclasses: correct and
incorrect vertex selection~\cite{nim}. An incorrect vertex can be
selected when two or more collisions occur in one beam bunch crossing,
making it possible that the highest reconstructed $\sum_{\rm
  tracks}p_{T}$ vertex does not produce the photon. While the fraction
of events with incorrect vertices depends on the final event selection
criteria, the $t_{\rm c}^{\gamma}$ distribution for each subclass is
estimated separately using \Wele\ data where the electron track is
dropped from the vertexing.  For events with a correctly associated
vertex, the $t_{c}^{\gamma}$ distribution is Gaussian and centered at
zero with a standard deviation of 0.64~ns~\cite{nim}. For those with
an incorrectly selected vertex the $t_{\rm c}^{\gamma}$ distribution
is also Gaussian with a standard deviation of 2.05~ns.


The $t_{c}^{\gamma}$ distributions for both non-collision backgrounds
are estimated separately from data using events with no reconstructed
tracks. Photon candidates from cosmic rays are not correlated in time
with collisions, and therefore their $t_{\rm c}^{\gamma}$ distribution
is roughly flat. Beam halo photon candidates are produced by muons
that originate upstream of the detector (from the $p$ direction) and
travel through the calorimeter, typically depositing small amounts of
energy. When the muon deposits significant energy in the EM
calorimeter, it can be misidentified as a photon and cause $\mett$.
These photons populate predominantly the negative $t_{\rm c}^{\gamma}$
region, but can contribute to the signal region. Since beam halo muons
travel parallel to the beam line, these events can be separated from
cosmic ray events by identifying the small energy deposited in the
calorimeter towers along the beam halo muon trajectory.
 

The background prediction uses control regions outside the signal time
window but well within the 132~ns time window that the calorimeter
uses to measure the energy. The non-collision background templates are
normalized to match the number of events in two time windows: a beam
halo-dominated window at $\{-20$,~$-6\}$~ns, selected to be 3$\sigma$
away from the wrong vertex collision background, and a cosmic
rays-dominated window at $\{25$,~$90\}$~ns, well away from the
standard model and beam halo contributions. The collision background
is estimated by fitting events in the $\{-10$,~$1.2\}$~ns window with
the non-collision contribution subtracted and with the fraction of
correct to incorrect vertex events allowed to vary. In this way the
background for the signal region is entirely estimated from data
samples. The systematic uncertainty on the background estimate is
dominated by our ability to calibrate the mean of the $t_{\rm
  c}^{\gamma}$ distribution for prompt photons.  We find a variation
of 200~ps on the mean and 20~ps on the standard deviation of the
distribution by considering various possible event selection criteria.
These contribute to the systematic uncertainty of the collision
background estimate in the signal region and are added in quadrature
with the statistical uncertainties of the final fit procedure.


We estimate the sensitivity to heavy, long-lived particles that decay
to photons using GMSB models for different \none\ masses and
lifetimes. Events from all SUSY processes are simulated with the {\sc
  pythia} Monte Carlo program~\cite{pythia} along with the detector
simulation~\cite{geant}. The acceptance is the ratio of simulated
events that pass all the requirements to all events produced. It is
used in the optimization procedure and in the final limit setting and
depends on a number of effects. The fraction of \neut\ decays in the
detector volume is the dominant effect on the acceptance. For a given
lifetime this depends on the boost of the \neut. A highly boosted
\neut\ that decays in the detector typically does not contribute to
the acceptance because it tends to produce a photon traveling in the
same direction as the \neut. Thus, the photon's arrival time is
indistinguishable from promptly produced photons. At small boosts the
decay is more likely to happen inside the detector, and the decay
angle is more likely to be large, which translates into a larger delay
for the photon. The fraction of events with a delayed photon arrival
time initially rises as a function of \neut\ lifetime, but falls as
the fraction of \none's decaying outside the detector begins to
dominates. In the \none\ mass region considered ($65\le
m_{\tilde{\chi}_1^0}\le 150$~\gevc), the acceptance peaks at a
lifetime of around 5~ns.  The acceptance also depends on the mass as
the boost effects are mitigated by the ability to produce high energy
photons or \met\ in the collision, as discussed in
Ref.~\cite{prospects}.


The total systematic uncertainty of 10\% on the acceptance is
dominated by the uncertainty on the mean of the $t_{c}^{\gamma}$
distribution (7\%) and on the photon ID efficiency (5\%). Other
significant contributions come from uncertainties on initial and final
state radiation (3\%), jet energy measurement
(3\%), 
and the parton distribution functions (1\%).


We determine the kinematic and $t_{c}^{\gamma}$ selection requirements
that define the final data sample by optimizing the expected cross
section limit without looking at the data in the signal region.  To
compute the expected 95\% confidence level (C.L.) cross section upper
limit~\cite{Boos}, we combine the predicted GMSB signal and background
estimates with the systematic uncertainties using a Bayesian method
with a flat prior~\cite{Conway}.  The expected limits are optimized by
simultaneously varying the selection requirements for \mett, photon
\Et, jet \Et, azimuth angle between the leading jet and \mett\
($\Delta\phi($\mett, jet)), and $t_{c}^{\gamma}$. The
$\Delta\phi($\mett, jet) requirement rejects events where the \mett\
is overestimated because of a poorly measured jet. While each point in
\neut\ lifetime vs. mass space gives a slightly different
optimization, we choose a single set of requirements because it
simplifies the final analysis, while only causing a small loss of
sensitivity. The optimized requirements are summarized in
Table~\ref{tab:select}. As an example, the acceptance for
$m_{\tilde{\chi}_1^0}$~=~100~\gevc\ and lifetime
$\tau_{\tilde{\chi}_1^0}$~=~5~ns is estimated to be (6.3$\pm$0.6)\%.

\begin{figure}[tbp]
  \centering%
  \includegraphics[width=1\linewidth]{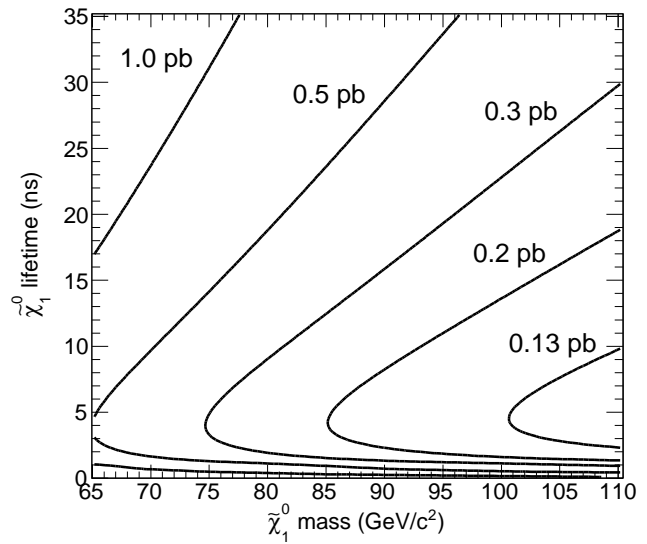}
  \caption{The contours of constant 95\% C.L. upper cross section
    limits for a GMSB model~\cite{snowmass}.}
  \label{fig:croscont}
\end{figure}

\begin{figure}[tbp]
  \centering%
  \includegraphics[width=1\linewidth]{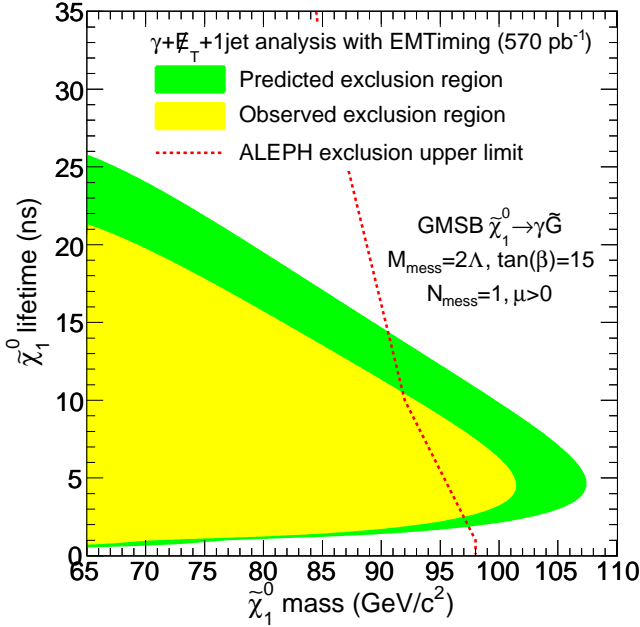}
  \caption{The exclusion region at 95\% C.L. as a function of \none\
    lifetime and mass for a GMSB model~\cite{snowmass}. The predicted
    and the observed regions are shown separately and are compared to
    the most stringent published limit from LEP searches~\cite{lep}.}
  \label{fig:exclreg}
\end{figure}

After all kinematic requirements, 508 events are observed in the data
before the final signal region time requirement. Their time
distribution is shown in Fig.~\ref{fig:data}. Our fit to the data
outside the signal region predicts total backgrounds of 6.2$\pm$3.5
from cosmic rays, 6.8$\pm$4.9 from beam halo background sources, and
the rest from the standard model. Inside the signal time region,
$\{2$, $10\}$~ns, we predict 1.25$\pm$0.66 events: 0.71$\pm$0.60 from
standard model, 0.46$\pm$0.26 from cosmic rays, and 0.07$\pm$0.05 from
beam halo. Two events are observed in the data. Since the result is
consistent with the no-signal hypothesis, we set limits on the \none\
lifetime and mass. Figure~\ref{fig:croscont} shows the contours of
constant 95\% C.L. cross section upper limit. Figure~\ref{fig:exclreg}
shows the exclusion region at 95\% C.L., along with the expected limit
for comparison.  This takes into account the predicted production
cross section at next-to-leading order~\cite{kfactors} as well as the
uncertainties on the parton distribution functions (6\%) and the
renormalization scale (2\%). Since the number of observed events is
above expectations, the observed limits are slightly worse than the
expected limits. These limits extend at large masses beyond those of
LEP searches using photon ``pointing'' methods~\cite{lep}.

In conclusion, we have performed the first search for heavy,
long-lived particles that decay to photons at a hadron collider using
data collected with the EMTiming system at the CDF~II detector. There
is no excess of events beyond expectations. As our search strategy
does not rely on event properties specific solely to GMSB models, we
can exclude any $\gamma$+jet+\mett\ signal that would produce more
than 5.5 events. We set cross section limits using a supersymmetric
model with \NTGTGR, and find a GMSB exclusion region in the \none\
lifetime vs. mass plane with the world-best 95\% C.L. lower limit on
the \none\ mass of 101~\gevc\ at $\tau_{\tilde{\chi}_1^0}$~=~5~ns.
Future improvements with similar techniques should also provide
sensitivity to new particle decays with a delayed electron
signature~\cite{feng}.  By the end of Run~II, an integrated luminosity
of 10~\invfb\ is possible for which we estimate a mass reach of
$\simeq 140$~\gevc\ at a lifetime of 5~ns.

\begin{acknowledgments}
  We thank the Fermilab staff and the technical staffs of the
  participating institutions for their vital contributions. This work
  was supported by the U.S. Department of Energy and National Science
  Foundation; the Italian Istituto Nazionale di Fisica Nucleare; the
  Ministry of Education, Culture, Sports, Science and Technology of
  Japan; the Natural Sciences and Engineering Research Council of
  Canada; the National Science Council of the Republic of China; the
  Swiss National Science Foundation; the A.P. Sloan Foundation; the
  Bundesministerium f\"ur Bildung und Forschung, Germany; the Korean
  Science and Engineering Foundation and the Korean Research
  Foundation; the Particle Physics and Astronomy Research Council and
  the Royal Society, UK; the Russian Foundation for Basic Research;
  the Comisi\'on Interministerial de Ciencia y Tecnolog\'{\i}a, Spain;
  in part by the European Community's Human Potential Programme under
  contract HPRN-CT-2002-00292; and the Academy of Finland.
\end{acknowledgments}

\end{document}

%% file: phys_defs.tex
\newcommand{\invpb}{$\rm pb^{-1}$} 
\newcommand{\invfb}{$\rm fb^{-1}$} 

\newcommand{\photon}{$\gamma$}
\newcommand{\ett}{\Et}
\newcommand{\ptt}{\mbox{$P_T$}}
\newcommand{\PT}{\ptt}
\newcommand{\NONE}{\mbox{$\widetilde{\chi}_1^0$}}
\newcommand{\NTWO}{\mbox{$\widetilde{\chi}_2^0$}}
\newcommand{\CONE}{\mbox{$\widetilde{\chi}_1^{\pm}$}}
\newcommand{\CONEP}{\mbox{$\widetilde{\chi}_1^{+}$}}
\newcommand{\CONEM}{\mbox{$\widetilde{\chi}_1^{-}$}}
\newcommand{\none}{\NONE}
\newcommand{\ntwo}{\NTWO}
\newcommand{\cone}{\CONE}
\newcommand{\coneb}{\CONEB}
\newcommand{\NTGNO}{\mbox{$\NTWO \rightarrow \gamma \NONE$}}
\newcommand{\NTGTGR}{\mbox{$\NONE \rightarrow \gamma \Gravitino$}}
\newcommand{\tanbeta}{\tan\beta}

\newcommand{\gevc}{GeV/$c^2$}
\newcommand{\punit}{GeV/$c$}

\newcommand{\mysect}[1]{\addtocounter{myctr}{1} \noindent \underline{Paragraph {\arabic{myctr}}: #1} \\}
\newcommand{\neut}{$ \widetilde{\chi}_1^0 $}
\newcommand{\grav}{\ensuremath{\widetilde{G}}}
\newcommand{\mett}{\mbox{$E\!\!\!\!/_{T}$}}
\newcommand{\met}{\mett}
\newcommand{\eeggmett}{$ee\gamma\gamma\mett $}
\newcommand{\ggmett}{$\gamma\gamma\mett $}
\newcommand{\gmett}{$\gamma$\met}
\newcommand{\scinotn}[2]{\mbox{${#1}\times 10^{#2}$}}
\newcommand{\sumet}{$\sum E_{T} $}
\newcommand{\sumpt}{$\sum P_{T} $}
\newcommand{\dphicosm}{$\Delta\phi $}
\newcommand{\chisq}{${\chi}^2 $}
\newcommand{\etal}{{\it et al.}}

\newcommand{\zpos}{ {\it Z} }
\newcommand{\trkz}{$Z_{0}$}
\newcommand{\trkt}{$T_{0}$}

\def\roots{{\sqrt s}}
\def \Et {$E_{T}$}
\def \Etg {E_{T}^{\gamma}}
\def\Z{{ Z^0}}
\def\selectron{\widetilde{e}}
\def\goes{\rightarrow}
\def\Gravitino{\widetilde{G}}
\def\pbarp{p{\bar p}}
\def\degrees{^\circ}
\def\deg{^\circ}
\def\ppbar{p{\bar p}}

\def\W{{ $W^{\pm}$}}
\def\Wele{ $W \rightarrow e\nu$}
\def\Zee{ $Z \rightarrow e^+e^-$}

\def\Journal#1#2#3#4{{#1}\textbf{~#2}, #4 (#3)}
\def\PrePrint#1{\mbox{hep-ph/#1}}
\def\PRL{\rm Phys. Rev. Lett.}
\def\PRD{{\rm Phys. Rev.} D}
\def\NIM#1#2#3{\rm Nucl. Instr. and Meth A{#1}~(#2)~#3}

\def\rphi{r-$\phi$}                                                                           
\def\rz{r-z}
\def\and{\\ \vspace{0.3cm}}

%% file: Jan_2007_Authors_Visitors.tex
\affiliation{Institute of Physics, Academia Sinica, Taipei, Taiwan 11529, Republic of China} 
\affiliation{Argonne National Laboratory, Argonne, Illinois 60439} 
\affiliation{Institut de Fisica d'Altes Energies, Universitat Autonoma de Barcelona, E-08193, Bellaterra (Barcelona), Spain} 
\affiliation{Baylor University, Waco, Texas  76798} 
\affiliation{Istituto Nazionale di Fisica Nucleare, University of Bologna, I-40127 Bologna, Italy} 
\affiliation{Brandeis University, Waltham, Massachusetts 02254} 
\affiliation{University of California, Davis, Davis, California  95616} 
\affiliation{University of California, Los Angeles, Los Angeles, California  90024} 
\affiliation{University of California, San Diego, La Jolla, California  92093} 
\affiliation{University of California, Santa Barbara, Santa Barbara, California 93106} 
\affiliation{Instituto de Fisica de Cantabria, CSIC-University of Cantabria, 39005 Santander, Spain} 
\affiliation{Carnegie Mellon University, Pittsburgh, PA  15213} 
\affiliation{Enrico Fermi Institute, University of Chicago, Chicago, Illinois 60637} 
\affiliation{Comenius University, 842 48 Bratislava, Slovakia; Institute of Experimental Physics, 040 01 Kosice, Slovakia} 
\affiliation{Joint Institute for Nuclear Research, RU-141980 Dubna, Russia} 
\affiliation{Duke University, Durham, North Carolina  27708} 
\affiliation{Fermi National Accelerator Laboratory, Batavia, Illinois 60510} 
\affiliation{University of Florida, Gainesville, Florida  32611} 
\affiliation{Laboratori Nazionali di Frascati, Istituto Nazionale di Fisica Nucleare, I-00044 Frascati, Italy} 
\affiliation{University of Geneva, CH-1211 Geneva 4, Switzerland} 
\affiliation{Glasgow University, Glasgow G12 8QQ, United Kingdom} 
\affiliation{Harvard University, Cambridge, Massachusetts 02138} 
\affiliation{Division of High Energy Physics, Department of Physics, University of Helsinki and Helsinki Institute of Physics, FIN-00014, Helsinki, Finland} 
\affiliation{University of Illinois, Urbana, Illinois 61801} 
\affiliation{The Johns Hopkins University, Baltimore, Maryland 21218} 
\affiliation{Institut f\"{u}r Experimentelle Kernphysik, Universit\"{a}t Karlsruhe, 76128 Karlsruhe, Germany} 
\affiliation{High Energy Accelerator Research Organization (KEK), Tsukuba, Ibaraki 305, Japan} 
\affiliation{Center for High Energy Physics: Kyungpook National University, Taegu 702-701, Korea; Seoul National University, Seoul 151-742, Korea; SungKyunKwan University, Suwon 440-746, Korea} 
\affiliation{Ernest Orlando Lawrence Berkeley National Laboratory, Berkeley, California 94720} 
\affiliation{University of Liverpool, Liverpool L69 7ZE, United Kingdom} 
\affiliation{University College London, London WC1E 6BT, United Kingdom} 
\affiliation{Centro de Investigaciones Energeticas Medioambientales y Tecnologicas, E-28040 Madrid, Spain} 
\affiliation{Massachusetts Institute of Technology, Cambridge, Massachusetts  02139} 
\affiliation{Institute of Particle Physics: McGill University, Montr\'{e}al, Canada H3A~2T8; and University of Toronto, Toronto, Canada M5S~1A7} 
\affiliation{University of Michigan, Ann Arbor, Michigan 48109} 
\affiliation{Michigan State University, East Lansing, Michigan  48824} 
\affiliation{University of New Mexico, Albuquerque, New Mexico 87131} 
\affiliation{Northwestern University, Evanston, Illinois  60208} 
\affiliation{The Ohio State University, Columbus, Ohio  43210} 
\affiliation{Okayama University, Okayama 700-8530, Japan} 
\affiliation{Osaka City University, Osaka 588, Japan} 
\affiliation{University of Oxford, Oxford OX1 3RH, United Kingdom} 
\affiliation{University of Padova, Istituto Nazionale di Fisica Nucleare, Sezione di Padova-Trento, I-35131 Padova, Italy} 
\affiliation{LPNHE, Universite Pierre et Marie Curie/IN2P3-CNRS, UMR7585, Paris, F-75252 France} 
\affiliation{University of Pennsylvania, Philadelphia, Pennsylvania 19104} 
\affiliation{Istituto Nazionale di Fisica Nucleare Pisa, Universities of Pisa, Siena and Scuola Normale Superiore, I-56127 Pisa, Italy} 
\affiliation{University of Pittsburgh, Pittsburgh, Pennsylvania 15260} 
\affiliation{Purdue University, West Lafayette, Indiana 47907} 
\affiliation{University of Rochester, Rochester, New York 14627} 
\affiliation{The Rockefeller University, New York, New York 10021} 
\affiliation{Istituto Nazionale di Fisica Nucleare, Sezione di Roma 1, University of Rome ``La Sapienza," I-00185 Roma, Italy} 
\affiliation{Rutgers University, Piscataway, New Jersey 08855} 
\affiliation{Texas A\&M University, College Station, Texas 77843} 
\affiliation{Istituto Nazionale di Fisica Nucleare, University of Trieste/\ Udine, Italy} 
\affiliation{University of Tsukuba, Tsukuba, Ibaraki 305, Japan} 
\affiliation{Tufts University, Medford, Massachusetts 02155} 
\affiliation{Waseda University, Tokyo 169, Japan} 
\affiliation{Wayne State University, Detroit, Michigan  48201} 
\affiliation{University of Wisconsin, Madison, Wisconsin 53706} 
\affiliation{Yale University, New Haven, Connecticut 06520} 
\author{A.~Abulencia}
\affiliation{University of Illinois, Urbana, Illinois 61801}
\author{J.~Adelman}
\affiliation{Enrico Fermi Institute, University of Chicago, Chicago, Illinois 60637}
\author{T.~Affolder}
\affiliation{University of California, Santa Barbara, Santa Barbara, California 93106}
\author{T.~Akimoto}
\affiliation{University of Tsukuba, Tsukuba, Ibaraki 305, Japan}
\author{M.G.~Albrow}
\affiliation{Fermi National Accelerator Laboratory, Batavia, Illinois 60510}
\author{S.~Amerio}
\affiliation{University of Padova, Istituto Nazionale di Fisica Nucleare, Sezione di Padova-Trento, I-35131 Padova, Italy}
\author{D.~Amidei}
\affiliation{University of Michigan, Ann Arbor, Michigan 48109}
\author{A.~Anastassov}
\affiliation{Rutgers University, Piscataway, New Jersey 08855}
\author{K.~Anikeev}
\affiliation{Fermi National Accelerator Laboratory, Batavia, Illinois 60510}
\author{A.~Annovi}
\affiliation{Laboratori Nazionali di Frascati, Istituto Nazionale di Fisica Nucleare, I-00044 Frascati, Italy}
% ...
\collaboration{CDF Collaboration\footnote{With visitors from $^a$University of Athens, 
$^b$University of Bristol, 
$^c$University Libre de Bruxelles, 
$^d$Cornell University, 
$^e$University of Cyprus, 
$^f$University of Dublin, 
$^g$University of Edinburgh, 
$^h$University of Heidelberg, 
$^i$Universidad Iberoamericana, 
$^j$University of Manchester, 
$^k$Nagasaki Institute of Applied Science, 
$^l$University de Oviedo, 
$^m$University of London, Queen Mary College, 
$^n$University of California Santa Cruz, 
$^o$Texas Tech University, 
$^p$University of California Irvine, 
$^q$IFIC(CSIC-Universitat de Valencia), 
}}
\noaffiliation

%% file: llived_prl.bbl
\begin{thebibliography}{99}
\bibitem{rapdef} We use a cylindrical coordinate system in which the
  proton beam travels along the $z$-axis, $\theta$ is the polar angle,
  $\phi$ is the azimuthal angle, and $\eta=-\ln \tan(\theta/2)$. The
  transverse energy and momentum are defined as $E_T=E\sin\theta$ and
  $p_T = p \sin \theta$ where $E$ is the energy measured by the
  calorimeter and $p$ the momentum measured in the tracking system.
  $\met=|-\sum_i E_T^i \vec{n_i}|$ where $\vec{n_i}$ is a unit vector
  that points from the interaction vertex to the $i$th calorimeter
  tower in the transverse plane.
\bibitem{zvtime} The distribution of the $\pbarp$ collisions has a
  standard deviation of 30~cm and 1.3~ns in $z_i$ and $t_i$,
  respectively.
\end{thebibliography}
